\documentclass[aps,prb,twocolumn,superscriptaddress,showpacs,amsmath,amssymb]{revtex4-1}

\usepackage{graphicx}
\usepackage{bm}

\begin{document}

\title{Anisotropic criteria for the type of superconductivity}

\author{V. G. Kogan}
\email{kogan@ameslab.gov}
\affiliation{Ames Laboratory - DOE and Department of Physics, Iowa State University, Ames, IA 50011}

\author{R. Prozorov}
\email{prozorov@ameslab.gov}
\affiliation{Ames Laboratory - DOE and Department of Physics, Iowa State University, Ames, IA 50011}

\begin{abstract}
The classical criterion for classification of superconductors as type-I or type-II based on the isotropic Ginzburg-Landau theory is generalized to arbitrary temperatures for materials with anisotropic Fermi surfaces and order parameters. We argue that the relevant quantity for this classification is the ratio of the upper and thermodynamic critical fields, $H_{c2}/H_c$, rather than the traditional ratio of the penetration depth and the coherence length, $\lambda/\xi$. Even in the isotropic case,  $H_{c2}/H_c$ coincides with $\sqrt{2}\lambda/\xi$ only at the critical temperature $T_c$ and they differ as $T$ decreases, the long known fact. Anisotropies of Fermi surfaces  and order parameters may amplify this difference  and render false the criterion based on the value of $\kappa=\lambda/\xi$.
\end{abstract}



\date{24 July 2014}

\maketitle \maketitle

\section{Introduction}

The classification of superconductors as type-I and type-II introduced within the Ginzburg-Landau (GL) theory near $T_c$  is based on the value of the GL parameter $\kappa=\lambda/\xi$ ($\lambda$ is the weak field penetration depth and $\xi$ is the coherence length).\cite{GL,Abrik}  A bulk material is of the type-II if $\kappa>1/\sqrt{2}$; in fields $H>H_{c1}\approx (\phi_0/4\pi\lambda^2)(\ln\kappa+0.5 )$ vortices are nucleated.\cite{ChiaRenHu} The lower critical  field $H_{c1}$ is  related to the line energy of a single vortex, $\varepsilon_l$, which is found by solving the GL equations for the order parameter and supercurrents: $H_{c1}=\phi_0 \varepsilon_l/8\pi$. The mixed phase with vortices exists in fields up to $H_{c2}= \phi_0/2\pi\xi^2 $ such that $H_{c1}<H_c<H_{c2}$, where the thermodynamic critical field  is related to the condensation energy density $F=H_c^2/8\pi$.  In the GL domain $H_c= \phi_0/2\sqrt{2}\pi\xi\lambda $.
  If $\kappa<1/\sqrt{2}$,  the bulk material is in the Meissner state in fields $H<H_c$ and   is classified as type-I.

The question of this classification for low temperatures in isotropic materials  was addressed by Eilenberger who evaluated the upper critical field $H_{c2}$  along with $H_c$
to show that $\kappa_1=H_{c2}(T)/\sqrt{2}H_c(T)$ increases on cooling to $T=0$ by about 30\%. \cite{Eil} Hence, taking $\kappa$ as governing material behavior in magnetic field, one concludes that if  $\kappa>1/\sqrt{2}$ at $T_c$, it certainly exceeds this value at all temperatures and, therefore, the GL classification should hold at any $T$. It is worth noting that this classification holds   for  Fermi  spheres and constant order parameters (s-wave).

When strongly anisotropic materials came forth and in particular with discovery of cuprates, it was realized that a mere fact of anisotropy may cause  $\lambda/\xi$ to  change with the field orientation.\cite{Buzdin} Although for cuprates with $\lambda\gg \xi$ the question of the superconductivity type never arose, it became clear  that in principle an anisotropic material can be type-I for one field orientation and type-II for another.

The situation is even more complicated  with multi-band materials and with other than s-wave order parameters for which the temperature and angular behavior of $H_{c2}$ (along with $\xi$) differs from that of  $\lambda$, while both these quantities depend on the Fermi surface and on the order parameter anisotropy.

The general formalism for calculating $H_{c2}$ and  $\lambda$ in the clean case has recently been developed for arbitrary Fermi surfaces and order parameters.\cite{K2002,PK-ROPP,KP-ROPP} We argue, however, that minute details of the Fermi surfaces are usually of  little effect on $H_{c2}$ and  $\lambda$ because the equations governing these quantities contain only integrals over the whole Fermi surfaces. Therefore, one  can consider the  simplest Fermi shapes of spheroids (for tetragonal materials) for which the Fermi surface averaging is a well defined procedure. Hence,   $\kappa(T)$ is now accessible for various  anisotropies of Fermi surfaces and order parameters.

However, for anisotropic materials at arbitrary temperatures, the GL  criterion based on the value of $\kappa=\lambda/\xi$ is questionable  because the GL theory {\it per se} only works near $T_c$. We use in this text a different approach based on the fact that in type-II superconductors
the two characteristic fields, $H_{c1}$  at which vortices nucleate in the bulk material, and $H_{c2}$, the maximum field at which the mixed state exists, satisfy   $H_{c1}< H_c< H_{c2}$. Either part of this inequality, $H_{c1}< H_c $ or $ H_c< H_{c2}$ (or for this matter $H_{c1}<  H_{c2}$), can be used to classify the material behavior as that of type-II. However, to have $H_{c1}(T)$ one should  evaluate    the vortex line energy within the microscopic theory, a difficult problem  if at all doable.  On the other hand, both  $ H_{c2}(T)$ and $H_c(T)$
can be evaluated for  anisotropic   Fermi surfaces and order parameters   at any temperature.  It is the criterion $H_c(T) < H_{c2}(T)$ that we study in this work.

Below we calculate the condensation energy for  anisotropic situation at arbitrary temperatures. Next, we review methods for evaluation of $H_{c2}$ and $\lambda$ and present numerical  results to show that the criterion based on the ratio $H_{c2}/H_c$ differs substantially from that employing $\lambda/\xi$.

\section{Condensation energy }

Perhaps, the simplest formally for our purpose is the approach based on the Eilenberger quasiclassical formalism
 that holds for a general anisotropic Fermi surface and for any gap symmetry. \cite{E} The theory deals with two functions, $f$ and $g$, which are integrated over the energy Gor'kov Green's functions.  For a uniform state of clean superconductors of interest here $f,g$ satisfy:
\begin{eqnarray}
  \Delta\, g - \hbar\omega f =0\,,\qquad
  g^2+f^2=1 \,.
\label{Eil12}
\end{eqnarray}
Here, $\hbar\omega=\pi T(2 n+1)$ with an integer $n$. We employ the approximation of a separable coupling responsible for superconductivity: $V(\bm k,\bm k^\prime)=V_0\Omega(\bm k)\Omega(\bm k^\prime)$, $\bm k$ is the Fermi momentum.\cite{Kad} In this approximation the order parameter $\Delta(T,\bm k)=\Psi(T)\Omega(\bm k)$.  $ \Omega({\bf k})$  determines the $\bm k$ dependence of $\Delta$ and is normalized so that the average over the Fermi surface
$\langle\Omega^2\rangle=1$.
  Equations (\ref{Eil12}) give:
\begin{eqnarray}
 f=  \Delta/ \beta\,,\quad g=   \omega/ \beta\,,\quad   \beta^2 =  \Delta^2+  \omega^2\,.
\label{f,g}
\end{eqnarray}
The order parameter should satisfy
  the self-consistency equation of the theory, see, e.g., Ref.\,\onlinecite{K2002}:
\begin{eqnarray}
 \frac{\Psi}{2\pi T} \ln \frac{T_{c}}{T}=\sum_{\omega> 0}
\left (\frac{\Psi}{ \hbar\omega}- \Big\langle
\Omega f\Big\rangle\right )\,,
\label{self-cons}
\end{eqnarray}
where $\langle...\rangle$ stands for averaging over the Fermi surface.

Equations (\ref{Eil12}) and (\ref{self-cons}) can be obtained as minimum conditions for  the energy functional:\cite{E}
\begin{eqnarray}
\frac{{\cal F}}{N(0) }=  \Psi^2  \ln \frac{T_{c}}{T} +2\pi T  \sum_{\omega> 0}
\left [\frac{\Psi^2}{\hbar \omega}- 2\Big\langle
 \Delta f+\hbar\omega (g-1)\Big\rangle\right ]\nonumber\\
\label{functional}
\end{eqnarray}
where $g=\sqrt{1-f^2}$ and $N(0)$ is the density of states per spin on the Fermi level. Substituting here the solutions (\ref{f,g}) and taking into account the self-consistency relation (\ref{self-cons}) one obtains the condensation energy density $F$:
\begin{eqnarray}
\frac{ F}{2\pi TN(0) }=  \left\langle \sum_{\omega> 0}
 \frac{(\beta-\hbar\omega)^2}{\beta}\right\rangle\,. \label{energy}
\end{eqnarray}

At $T=0$ (replace $2\pi T\sum_\omega \to \int_0^\infty  \hbar\, d\omega $),
 \begin{eqnarray}
F(0)=\frac{N(0)}{2}\langle\Delta^2(0)\rangle = \frac{N(0)}{2}\Psi^2(0)
\label{F(0)}
\end{eqnarray}
(recall the isotropic result $F(0)=N(0)\Delta^2(0)/2$). To find the value of $\Psi(0)$ one considers the first sum in Eq.\,(\ref{self-cons}) as extended to  $n_{\rm max} =\hbar\omega_D/2\pi T$, while the second is replaced with $ \int_0^{\hbar\omega_D}d( \hbar\omega)/2\pi T$ ($\omega_D$ is the Debye frequency for the phonon mechanism or a proper  cutoff for others):
\begin{eqnarray}
  \ln \frac{T_{c}}{T}=\ln\frac{2 e^{\bm C}\hbar\omega_D}{\pi T} -
\left \langle
\Omega^2 \ln\frac{2\hbar\omega_D}{\Psi |\Omega|}\right\rangle \,,
\label{self-cons0}
\end{eqnarray}
where $\bm C\approx 0.577$ is the Euler constant. This gives:
\begin{eqnarray}
    \Psi(0)  = \frac{\pi T_c} { e^{ \bm C}}\,e^{-\langle\Omega^2\ln|\Omega |\rangle}.
\label{Psi(0)}
\end{eqnarray}
Hence, we have $H_c(0)= 2\sqrt{\pi N(0)}\,\Psi(0)$.

Near $T_c$, Eq.\,(\ref{self-cons}) yields
\begin{eqnarray}
    \Psi^2   = \frac{8\pi^2 T_c^2 (1-t)} { 7\zeta (3)\langle\Omega^4 \rangle},
\label{Psi(Tc)}
\end{eqnarray}
where $t=T/T_c$. The condensation energy is readily found:
\begin{eqnarray}
   F   = \frac{7\zeta (3)N(0)  \langle\Omega^4 \rangle\Psi^4}{16\pi^2T_c^2}= \frac{4\pi^2T_c^2N(0)}{7\zeta (3)\langle\Omega^4 \rangle}(1-t)^2. \qquad
\label{Psi(Tc)}
\end{eqnarray}
Given   $F(T)$, it is straightforward to obtain the difference of specific heats $C_s-C_n$ at any $T$ and in particular the specific heat jump at $T_c$:\cite{Nagi,Openov}
\begin{eqnarray}
   \frac{\Delta C}{C_n}  = \frac{12}{7\zeta (3)  \langle\Omega^4 \rangle } = \frac{1.43}{  \langle\Omega^4 \rangle } \,.
\label{DC}
\end{eqnarray}
Near $T_c$, we have
 \begin{eqnarray}
H_c=   8\pi T_c \sqrt{\frac{\pi N(0)}{14\zeta(3)\langle  \Omega^4 \rangle}}  \,(1-t)\,.
  \label{Hc(Tc)}
\end{eqnarray}

For the numerical work at arbitrary temperatures, we rewrite the energy as
  \begin{eqnarray}
 F&=&4\pi^2T_c^2N(0)\, t^2S\,, \nonumber\\
 S&=&\sum_{n=0}^\infty\left\langle \frac{\left[\sqrt{(n+1/2)^2+\psi^2\Omega^2}-(n+1/2)\right]^2}{\sqrt{(n+1/2)^2+\psi^2\Omega^2}}\right\rangle \qquad
  \label{energy-num}
\end{eqnarray}
where  $\psi=\Psi/2\pi T$.
Thus, the general scheme of evaluation of the thermodynamic critical field consists of solving the self-consistency equation (\ref{self-cons}) for $\Psi(T)$ at each $T$ and then evaluating $F$
 of Eq.\,(\ref{energy-num}) and $H_c= \sqrt{8\pi F}$.

 As mentioned in Introduction, describing  Fermi surface shapes within problems of $H_c$ and $H_{c2}$,  one can consider   Fermi ellipsoids, for which the averaging is a well defined analytic procedure.\cite{MMK,KP-ROPP}  Although straightforward, this procedure is quite involved, a  brief description is given in Appendix A.

   Hence we  characterize  Fermi surfaces for tetragonal materials by a single parameter $\epsilon$,  the squared ratio of the spheroid semi-axes. We consider only representative  order parameters: s-wave ($\Omega=1$), d-wave  ($\Omega=\sqrt{2}\cos 2\varphi$ with $ \varphi$ being the azimuth of spherical coordinates with the polar axis along the $c$  crystal direction, and order parameters of the form   $\Omega=\Omega_0\cos^n \theta$ with the polar angle $\theta$. The latter were recently suggested as possibilities for at least some of the Fe-based materials;\cite{theory-nodes,arXiv1} the ``equatorial" node $n=1$ has also been observed in the ARPES data  on BaFe$_2$(As$_{0.7}$P$_{0.3}$)$_2$. \cite{Feng}
  \begin{figure}[t]
\begin{center}
\includegraphics[width=8.cm] {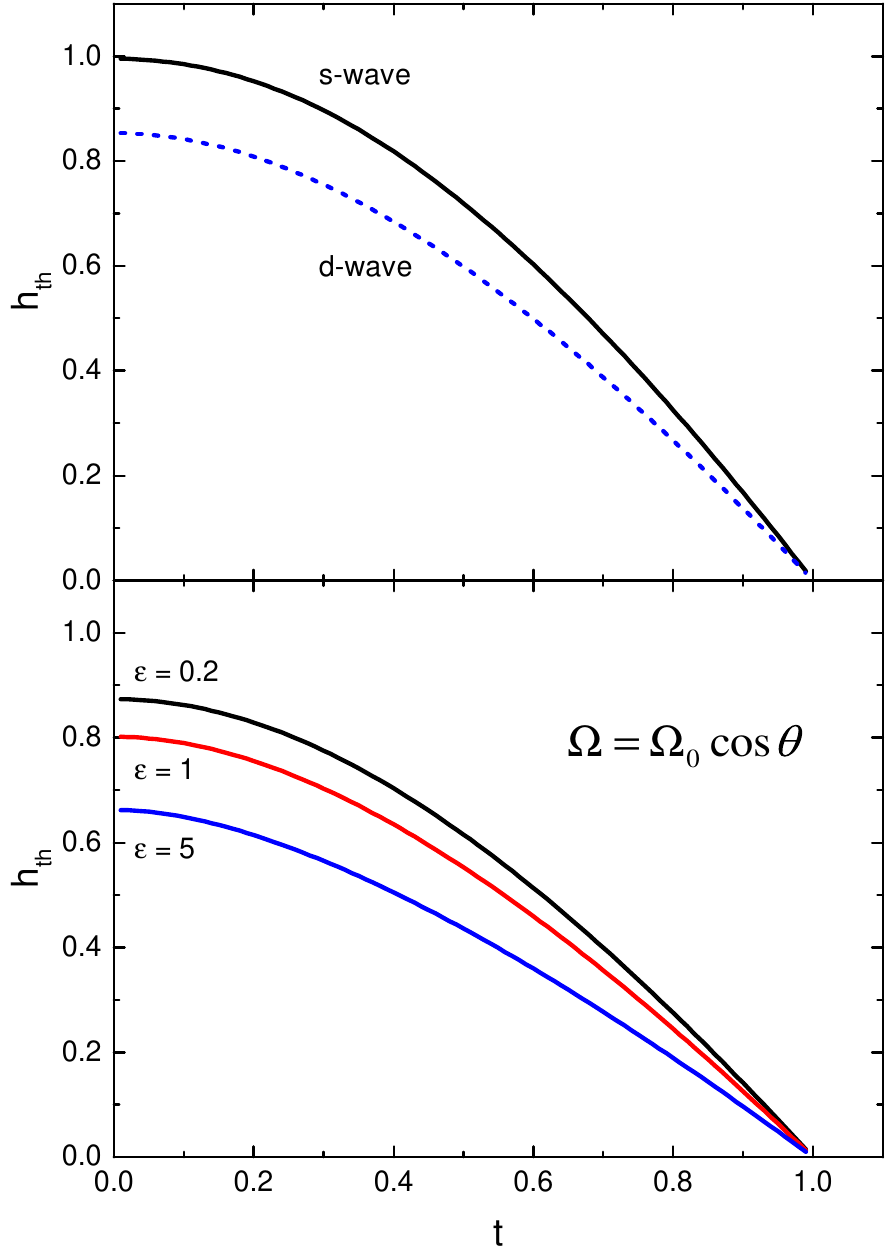}
\caption{(Color online) Dimensionless thermodynamic critical field $h_{th}(t)=H_c/ 2\pi T_c\sqrt{N(0)}$. Each curve on the upper panel in fact is three coinciding curves for  Fermi sphere and prolate and oblate spheroids, $\epsilon=1, 0.2,$ and 5.  The lower panel is for the order parameter $\Omega \propto \cos\theta$ with the normalization $\Omega_0$ evaluated separately for each Fermi shape, see Appendix A.  }
\label{f1}
\end{center}
\end{figure}

Numerical results for  the thermodynamic critical field $H_c$ in units of $2\pi T_c\sqrt{N(0)}$ are shown in Fig.\,\ref{f1}. This normalization is chosen because for the s-wave order parameter on a sphere we have close to 1 value of
 \begin{eqnarray}
  h_{th}(0)=\frac{H_c(0)}{2\pi T_c\sqrt{N(0)}} = \frac{\sqrt{\pi}}{e^{\bm C}}\approx 0.995\,
   \label{ht(0)}
\end{eqnarray}
(the notation $h_{th}$ for the normalized $H_c$ is   to avoid confusion with the $c$ direction).
  As  is seen in Fig.\,\ref{f1},  nodes  suppress  the condensation energy   and $H_c$.   Besides, we observe that while the shape of the Fermi surface does not affect $H_c$ for s- and d-wave order parameters, the equatorial node clearly makes a difference.

  \section{Upper critical field}

 The theory of the orbital $H_{c2}$  of clean superconductors has recently been developed by the authors for arbitrary anisotropies of Fermi surfaces and order parameters.\cite{KP-ROPP} Within this theory, $H_{c2}^{(c)}$ along the $c$ axis of uniaxial crystals is found by solving an equation:
 \begin{eqnarray}
&&   \ln t= 2 h^{(c)} \int_0^{\infty}s\, \ln\tanh (st) \,
 \left \langle\Omega^2 \mu_c  e^{-\mu_c h^{(c)} s^2 }\right\rangle ds\,, \qquad
  \label{eq-hc}\\
&& h^{(c)}=H_{c2}^{(c)}\frac{\hbar^2v_0^2}{2\pi \phi_0T_c^2},\,\, \mu_c=\frac{v_x^2+v_y^2}{v_0^2} ,\,\, v_{0 }^3= \frac{2E_F^2}{\pi^2\hbar^3N(0) }  .\qquad
 \label{mu_c}
\end{eqnarray}
Here, $v_x,v_y$ are Fermi velocities in the $a,b$ plane,  $E_F$ is the Fermi energy,    the velocity $v_0=v_F$ for the isotropic case.   Hence, both   $\mu_c$ depending on the Fermi surface and $\Omega$ describing the order parameter anisotropy, enter the equation for $h^{(c)}$  under the integral over the Fermi surface. This is the reason why the simple spheroid with the shape   fixed by a single parameter, the ratio of semi-axes, suffices to describe major features of quantities of interest here.

The theory of Ref.\,\onlinecite{KP-ROPP} allows one to evaluate also the anisotropy parameter $\gamma_H=H_{c2}^{(a)}/H_{c2}^{(c)}$. Given $h^{(c)}(t)$, one solves Eq.\,(\ref{eq-hc}) in which $\mu_c$ is replaced with
$\mu_a=(v_x^2+\gamma_H^2v_z^2)/v_0^2$.

 In general, Eq.\,(\ref{eq-hc}) can be solved numerically, but if $T=0$ or $T\to T_c$,  the solutions are exact: \cite{KP-ROPP}
\begin{eqnarray}
&& h^{(c)}(0)=\exp( -\bm C -  \langle\Omega^2 \ln \mu_c   \rangle),\nonumber\\
&&   h^{(c)}(t\to 1)= \frac{8(1-t)}{7\zeta(3)  \left\langle \Omega^2\mu_c \right\rangle } \,.
    \label{hc(Tc)}
 \end{eqnarray}
For the isotropic case with $\Omega=1$ and $\langle \mu_c\rangle=2/3$, one reproduces the  Helfand-Werthamer   clean limit results.\cite{HW}

  After simple algebra we obtain:
 \begin{eqnarray}
\frac{H_{c2}^{(c)}(0)}{H_c(0)}&=&   \frac{  \phi_0 T_c  }{\hbar^2v_0^2    \sqrt{ \pi N(0)}} \exp\left  \langle\Omega^2\ln \frac{|\Omega |}{\mu_c} \right\rangle \,,
  \label{Hc2/Hc-0} \\
 \frac{H_{c2}^{(c)}}{H_c}\Big |_{T_c}&=&   \frac{2\sqrt{2}  \phi_0 T_c }{\hbar^2v_0^2  \sqrt{7\zeta(3)\pi N(0)}} \,\frac{\sqrt{\langle  \Omega^4 \rangle}}{ \left\langle \Omega^2\mu_c \right\rangle}.
  \label{Hc2/Hc-1}
\end{eqnarray}
In the isotropic case near $T_c$, $H_{c2} /H_c=\sqrt{2}\,\kappa_{GL}$ with
 \begin{eqnarray}
\kappa_{GL}=  \frac{3 \phi_0 T_c  }{ \hbar^2v_F^2    \sqrt{7\zeta(3)\pi N(0)}}  \,,\qquad
  \label{kapGL}
\end{eqnarray}
see  Refs.\,\onlinecite{Gorkov} or \onlinecite{Parks}; this coincides with the isotropic limit of Eq.\,(\ref{Hc2/Hc-1}).

As mentioned above, if the ratio $R=H_{c2}/ H_c>1$, the material in question is of the type-II, if $R<1$ it behaves as type-I. Using Eqs.\,(\ref{Hc2/Hc-0}) and (\ref{Hc2/Hc-1}) we  compare these ratios at $T=0$ and $T_c$ for the $c$ direction:
 \begin{eqnarray}
\frac{R^{(c)}(0)}{R^{(c)}(T_c) }=  \sqrt{ \frac{  7\zeta(3) }{8}}  \,\frac{\langle  \Omega^2\mu_c\rangle}{\sqrt{\langle  \Omega^4 \rangle} } \exp\left  \langle\Omega^2\ln \frac{|\Omega |}{\mu_c} \right\rangle \,.
  \label{R(0)/R(1)}
\end{eqnarray}
It is worth noting that this ratio depends on the Fermi surface shape and the order parameter symmetry, but not on other material characteristics.

As an example we take $\Omega=\sqrt{3}\cos\theta$ on a Fermi sphere to obtain $R^{(c)}(0)/R^{(c)}(T_c)\approx 1.365$. We  note again that for the same order parameter anisotropy, say, for $\Omega=\Omega_0\cos\theta$, the normalization $\langle\Omega^2\rangle=1$ imposes different $\Omega_0$ for different Fermi surfaces, see Appendix A and Fig.\,\ref{f7}.
Hence, the criteria for type-I or -II behavior depend  on the Fermi surface shape and the order parameter symmetry.

\section{Penetration depth}

   The  inverse tensor of squared penetration depth for the general anisotropic clean case is: \cite{K2002,PK-ROPP}
\begin{equation}
(\lambda^2)_{ik}^{-1}= \frac{16\pi^2 e^2 N(0) T}{c^2}\, \sum_{\omega>0} \Big\langle\frac{
\Delta^2v_iv_k}{\beta ^{3}}\Big\rangle \,.
\label{lambda-tensor}
\end{equation}
Here $\Delta=\Psi \Omega $, $\beta=\sqrt{\Delta^2+\hbar^2\omega^2}$, and $\Psi(T)$ satisfies the self-consistency equation:
 \begin{eqnarray}
     - \ln t =\sum_{n=0}^\infty
\left (\frac{1}{n+1/2}-\left\langle\frac{ \Omega^2}{\sqrt{\psi^2\Omega^2+(n+1/2)^2}}\right\rangle\right) \,\qquad
\label{psi}
\end{eqnarray}
where $\psi=\Psi/2\pi T$.

The  density  of states $N(0)$, Fermi velocities $\bm v$, and the order parameter anisotropy    $\Omega$ are the input parameters for
evaluation of $\lambda_{aa}$ and $\lambda_{cc}$.
$N(0)$ is not needed if one is interested only in the anisotropy $\gamma_\lambda=\lambda_{cc}/\lambda_{aa}$:
\begin{eqnarray}
\gamma^2_\lambda&=& \frac{\lambda_{aa}^{-2}}{\lambda_{cc}^{-2}}
=\frac{ \sum_n \left\langle
 \Omega ^2v_a^2/\eta^{3/2} \right\rangle}{  \sum_n \left\langle
 \Omega ^2v_c^2/\eta^{3/2} \right\rangle}\,, \nonumber\\
\eta&=&\psi^2\Omega^2+(n+1/2)^2\,.
  \label{gam-lam}
\end{eqnarray}
   It is easy to show that   Eq.\,(\ref{gam-lam}) gives:\cite{Gork,K2002}
 \begin{eqnarray}
\gamma^2_\lambda(0) =\frac{  \langle  v_a^2\rangle} { \langle  v_c^2\rangle} \,,   \qquad\gamma^2_\lambda(T_c) =\frac{  \langle\Omega^2 v_a^2\rangle}
 { \langle\Omega^2 v_c^2\rangle}\,.
  \label{0,Tc}
\end{eqnarray}

At first sight, $\gamma_\lambda$ should approach $T_c$ as a constant or at least as some power  $(1-t)^p$ with $p>1$.  This would mean that $\gamma_\lambda\approx {\rm const}$ in a practically  finite GL domain. This, however,  is not the case. To see this we
  evaluate   $\gamma_\lambda$ near $T_c$ where
 \begin{eqnarray}
 \eta^{3/2}=(n+1/2)^3\left(1+\frac{3\psi^2\Omega^2}{2(n+1/2)^2}\right)
  \label{eta}
\end{eqnarray}
since $\psi^2\ll 1$. Expanding Eq.\,(\ref{gam-lam}) for $\gamma_\lambda$ in powers of $\psi^2$ we obtain the first correction:
 \begin{eqnarray}
\gamma_\lambda= \gamma_\lambda(T_c) -
  \frac{93\,\zeta(5)}{28\,\zeta(3)}\left(\frac{\langle\Omega^4 v_a^2\rangle}{ \langle\Omega^2 v_a^2\rangle}-\frac{\langle\Omega^4 v_c^2\rangle}{ \langle\Omega^2 v_c^2\rangle}\right)  \psi^2\,.\qquad
\label{C}
 \end{eqnarray}
Since $\psi^2\propto (1-t)$, $\gamma_\lambda$ approaches $T_c$ with a non-zero slope  for  all order parameters except the s-wave with $\Omega=1$.

We will see below that for  general anisotropies the ratios $H_{c2}/H_c$ and $\lambda/\xi$ also attain their GL values  only at $T_c$ approaching them with finite slopes.\cite{arXiv}

\section{Isotropic case}

This well-studied case is worth recalling because already here one can see that the criterion based on the value of $\lambda/\xi$ cannot be applied at arbitrary temperatures.
We obtain using Eq.\,(\ref{R(0)/R(1)}):
 \begin{eqnarray}
\frac{R (0)}{R (T_c) }=  \sqrt{ \frac{  7\zeta(3) }{8}}  \,e^{2-\ln 4} \approx 1.263 \,,
  \label{R(0)/R(1)is}
\end{eqnarray}
  the value originally obtained by Eilenberger. \cite{Eil}
We thus see that if at $T_c$ an isotropic material has   $R(T_c)=\sqrt{2}\kappa_{GL}=1$  at the boundary between type-I and type-II,   it is of the type-II at $T=0$.  For the material to be of the type-I at all $T$'s, i.e., to have $R(t)<1$ at all temperature,   one needs $R(T_c)<1/1.263=0.792$, or $ \kappa_{GL}<0.792/\sqrt{2}=0.560$.  Moreover,   if  $0.560 <\kappa_{GL}<1/\sqrt{2}=0.707$   at $T_c$, the material should undergo the transition from type-I to type-II at some temperature under $T_c$.

It is easy to see that at all temperatures $T\ne T_c$ the criterion based on the ratio $H_{c2} /H_c$ differs from that based on   $\kappa=\lambda/\xi$. To this end we take microscopically calculated values at $T=0$:
 \begin{eqnarray}
&&\lambda^{-2}(0)=   \frac{8\pi e^2N(0)v_F^2} {3c^2}\,,\nonumber\\
&&\xi^2(0)=\frac{  \phi_0   }{2 \pi H_{c2}(0)  } =  \frac{ \hbar^2v_F^2  }{  \pi^2T_c^2   } e^{  \bm C -  2 }\,,
  \label{lam-xi-0}
   \end{eqnarray}
which give
\begin{eqnarray}
 \kappa^2(0)=   \frac{3\pi c^2T_c^2} {8e^2N(0)\hbar^2v_F^4} \,e^{ 2- {\bm C}  }\,.
  \label{kap(0)}
   \end{eqnarray}
Using the GL value for $\kappa(T_c)$ (\ref{kapGL}) we obtain:\cite{Eil}
\begin{eqnarray}
 \frac{\kappa (0)}{\kappa(T_c)}=  \sqrt{ \frac{7\zeta(3)} {24} \,e^{ 2- {\bm C} } }=1.206\,,
  \label{kap(0)/kapGl}
   \end{eqnarray}
This  differs from $R(0)/R(T_c)=1.263$ obtained above using the $H_{c2}/H_c$ criterion. The difference is not large, still it shows that even in the isotropic case the value of $\kappa=\lambda/\xi$ is not a correct criterion for the type of superconductivity at any temperature except $T_c$. Basically, this is because $H_{c2}/H_c=\kappa\sqrt{2}$ only at $T_c$.

  \begin{figure}[h]
\begin{center}
 \includegraphics[width=8.cm] {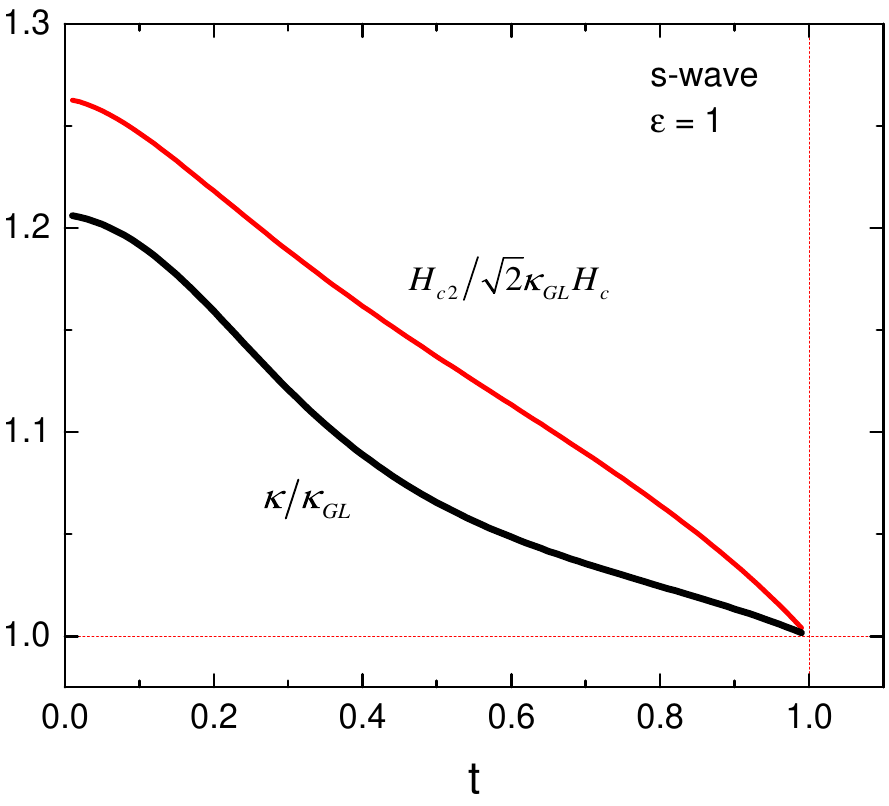}
\caption{(Color online) The red curve shows $ H_{c2}(t)/\sqrt{2}\kappa_{GL}H_c(t) $ and the lower curve is $\kappa(t)/\kappa_{GL}$ for the isotropic case.
 }
\label{f2}
\end{center}
\end{figure}

 These arguments are supported by the numerical calculation at arbitrary temperatures shown in Fig.\,\ref{f2}, where
the upper curve  is the ratio  $R(t)=H_{c2}(t)/H_c(t)$ for $\kappa_{GL}=1/\sqrt{2}$; the lower curve is $\kappa(t)/\kappa_{GL}$.
A feature worth noting in this figure is that the two curves have {\it finite} and {\it different} slopes at $T_c$. In other words,  in fact there is no however small temperature interval in the immediate vicinity of $T_c$ in which the GL ``$\kappa$-criterion" works, except $T_c$ itself.

This feature is related to the mentioned above accuracy of GL theory: the energy expansion within GL is accurate up to  terms of the order $\tau^2$ with $\tau=1-t$, the order parameter $\Psi^2\sim\tau$ along with   $\lambda^{-2}$,  $H_{c2}$, and $H_c$, all $\sim\tau$. Their ratios - within the GL theory - should be considered as constant. To get next corrections to these constants one has to overstep the accuracy of the GL theory, i.e., to go to the microscopic theory which   shows that these ratios approach $T_c$ with finite slopes.

\section{Numerical results}

The situation for anisotropic materials is, of course, more involved.  To begin,   we recall the standard notation. Introducing the geometric average  $\lambda=(\lambda_a^2\lambda_c)^{1/3} $ and  $\gamma_\lambda=\lambda_c/ \lambda_a$ one obtains  $\lambda_a= \lambda\gamma_\lambda^{-1/3}$ and $\lambda_c= \lambda\gamma_\lambda^{2/3}$ (for brevity we use the notation $\lambda_a$ instead of $\lambda_{aa}$ for the square root of one of diagonal  elements of  the tensor $(\lambda^2)_{ik}$). For the coherence lengths we have $\xi_a= \xi\gamma_H^{1/3}$ and $\xi_c= \xi\gamma_H^{-2/3}$, where $\gamma_H=H_{c2}^{(a)}/H_{c2}^{(c)}=\xi_a/\xi_c$ and $\xi^3=\xi_a^2\xi_c$. In general, $\gamma_H(T)\ne \gamma_\lambda(T)$, but at $T_c$ the anisotropies of both $\lambda$ and $H_{c2}$ are determined by the same ``mass tensor" so that $\gamma_H(T_c)=\gamma_\lambda(T_c)$.\cite{Gork,MMK,Carrington,arXiv} Different $\gamma_H(T)$ and $ \gamma_\lambda(T)$ demonstrate particularly well the common but misleading association  of  superconducting anisotropies with the effective mass tensor of the band theory.

Direct calculations of the thermodynamic critical field $H_c(T_c)$, either using the microscopic theory or the anisotropic GL equations, yield
\begin{eqnarray}
H_c(T_c)=\frac{\phi_0}{2 \sqrt{2} \pi \lambda_a\xi_a} =\frac{\phi_0}{2 \sqrt{2} \pi \lambda_c\xi_c}=\frac{\phi_0}{2 \sqrt{2} \pi \lambda \xi }  \,.
 \label{Hc-ac}
\end{eqnarray}
Hence, we have:
 \begin{eqnarray}
\frac{H_{c2}^{(c)}}{H_c}\Big |_{T_c}=\sqrt{2} \, \frac{\lambda\xi} {\xi_a^2}=\sqrt{2} \, \frac{\lambda_a} {\xi_a} =\sqrt{2} \kappa_a\,.\qquad
  \label{kap_a}
   \end{eqnarray}
   because $\gamma_\lambda/\gamma_H=1$ at $T_c$.
Using known $\lambda_a$ and $\xi_a$ we obtain skipping  the algebra:
 \begin{eqnarray}
\kappa_a=  \frac{ \phi_0 T_c  }{ \hbar^2v_0 }   \sqrt{\frac{2 \langle  \Omega^4 \rangle}{ 7\zeta(3)\pi N(0) \langle \Omega^2v^2_a \rangle\langle \Omega^2\mu_c  \rangle}}  \,.\qquad
  \label{kap-a}
\end{eqnarray}
It is easily verified that $\kappa_a$ reduces $\kappa_{GL}$ of Eq.\,(\ref{kapGL}) in the isotropic case.

For the in-plane field we have:
 \begin{eqnarray}
\frac{H_{c2}^{(a)}}{H_c}\Big |_{T_c}=\sqrt{2}   \frac{\lambda\xi} {\xi_a\xi_c} = \sqrt{2}   \frac{\lambda_c } {\xi_a } =  \sqrt{2} \,\kappa_\parallel.
  \label{kap_par}
   \end{eqnarray}
  Hence, for this field orientation, one should operate with parameter $\kappa_\parallel=\lambda_c /\xi_a $. This choice is also dictated by  the surface energy of the S-N boundary, say, in $(c,b)$ plane in field along $b$; the screening currents flow along $c$ whereas the order parameter is changing along $a$. Thus the relevant lengths in this case are $\lambda_c$ and $\xi_a$. We obtain:
 \begin{eqnarray}
\kappa_\parallel&=& \frac{\lambda_c}{\lambda_a}= \gamma_\lambda \kappa_a=\sqrt{\frac{\langle \Omega^2v^2_a \rangle}{\langle \Omega^2v^2_c \rangle}}\,\kappa_a\nonumber\\
&=&\frac{ \phi_0 T_c  }{ \hbar^2v_0 }   \sqrt{\frac{2 \langle  \Omega^4 \rangle}{ 7\zeta(3)\pi N(0) \langle \Omega^2v^2_c \rangle \langle \Omega^2\mu_c  \rangle}}  \,.\qquad
  \label{kap-par}
\end{eqnarray}

  For an arbitrary $T$, we obtain:
 \begin{eqnarray}
\frac{H_{c2}^{(c)}}{H_c}&=&   \frac{h^{(c)}(t)}{h_{th}} \,\frac{  \phi_0 T_c }{ \hbar^2v_0^2    \sqrt{N(0)}}  \,,\qquad
  \label{Hc2(t)/Hc(t)}\\
\frac{H_{c2}^{(a)}}{H_c}&=&   \frac{h^{(a)}(t)}{h_{th}} \,\frac{  \phi_0 T_c }{ \hbar^2v_0^2    \sqrt{N(0)}}   \,.
  \label{Hc2(t)/Hc(t)}
   \end{eqnarray}
  Presenting the numerical results we normalize the ratio $R^{(c)}=H_{c2}^{(c)}/ H_c$ to its value at $T_c$, i.e., to $\sqrt{2}\kappa_a$ whereas for the in-plane direction $R^{(a)}=H_{c2}^{(a)}/ H_c$ is normalized to $\sqrt{2}\kappa_\parallel$.

\begin{figure}[h]
 \begin{center}
  \includegraphics[width=7.5cm] {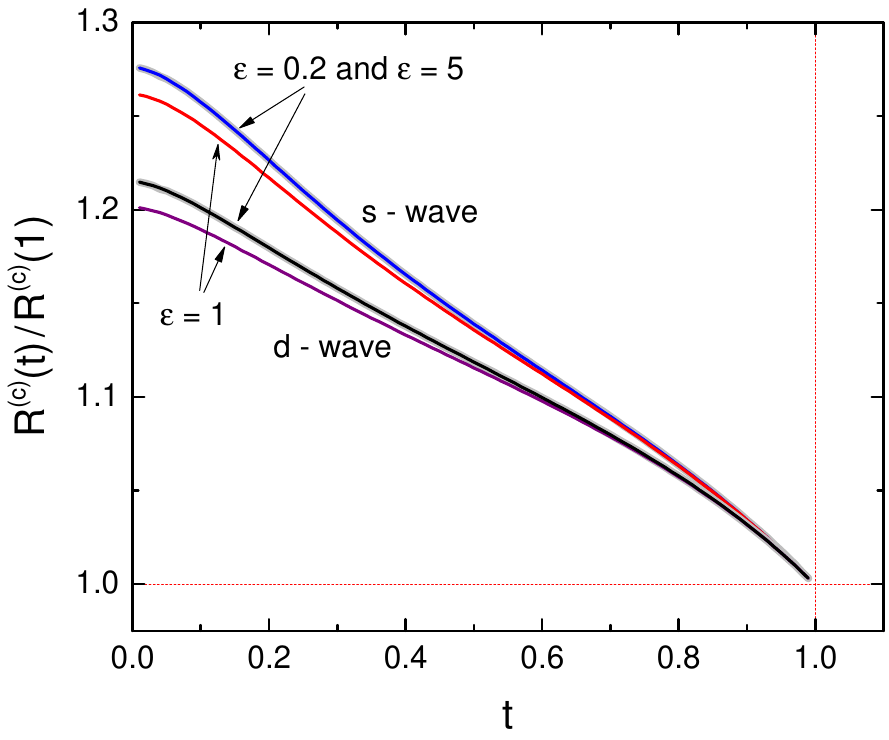}
 \caption{(Color online) The ratio $R(t)/R(1)$ for s- and d-waves; $R(t) =H_{c2}(t)/H_c(t)$.  Although the effect of the Fermi surface anisotropy is  weak, in both cases it results in increasing ratio of $H_{c2}/H_c$ at low-$T$'s
  }
 \label{f3}
 \end{center}
 \end{figure}

\begin{figure}[htb]
\begin{center}
 \includegraphics[width=7cm] {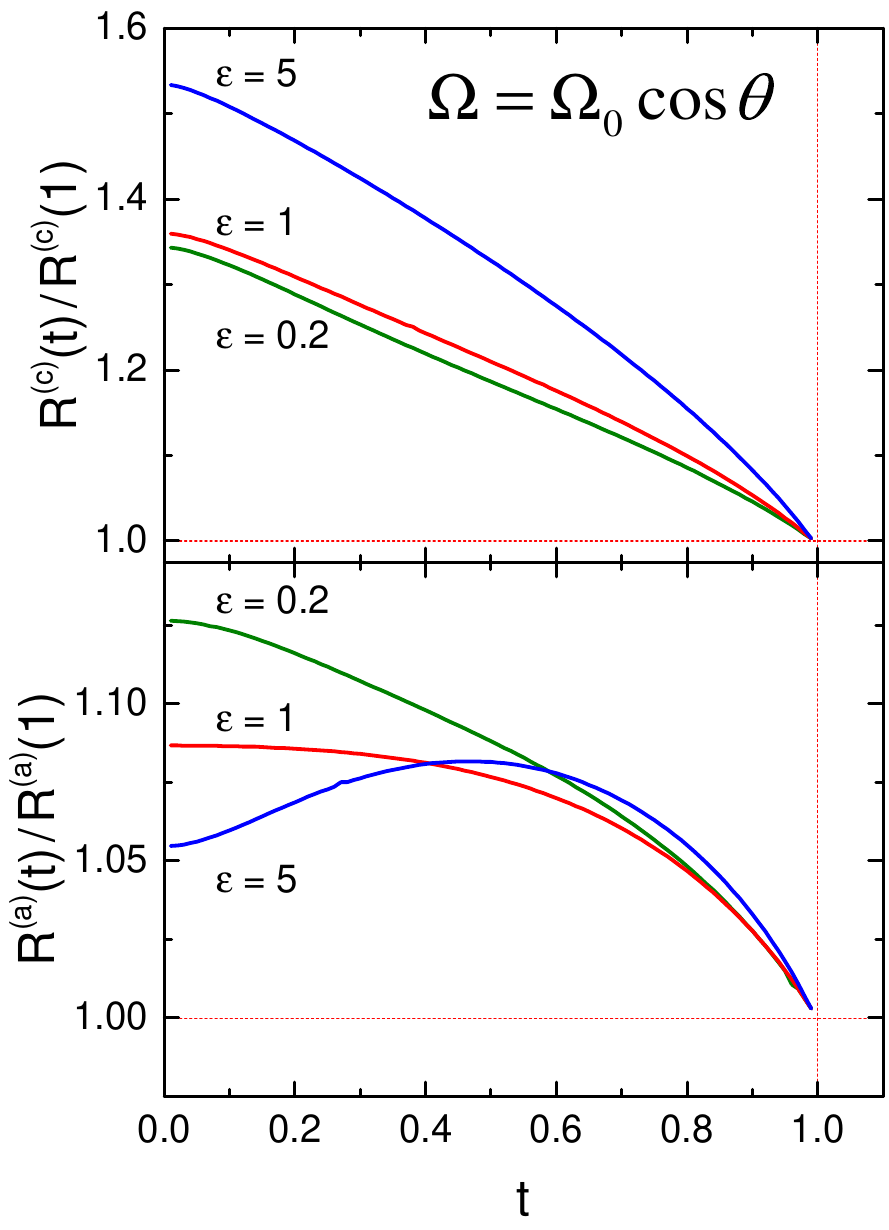}
\caption{(Color online) The ratio $R(t)/R(1)$ for two principal directions and three Fermi surface shapes:  prolate spheroid $\epsilon=0.2$, sphere, and oblate spheroid $\epsilon=5$. The order parameter has an equatorial node, $\Omega=\Omega_0\cos\theta$.
 }
\label{f4}
\end{center}
\end{figure}

Figure \ref{f3}  shows these normalized ratios for s- and d-wave order parameters, whereas
Fig.\,\ref{f4} is for the order parameter with an an equatorial node, $\Omega=\Omega_0\cos\theta$, for three Fermi surfaces: prolate spheroid $\epsilon=0.2$, sphere, and oblate spheroid $\epsilon=5$.  Note that $R^{(a)}(t) $ increases on cooling slower than $R^{(c)}(t) $ and can even go through a maximum   as it is in the oblate case of $\epsilon=5$. This behavior is related to the fact that  $R^{(a)} =\gamma_H  R^{(c)} $ and $\gamma_H(t)$  decreases on cooling for this order parameter, see Ref.\,\onlinecite{arXiv} and references therein.
One should bear in mind that for determining the material type  at a particular  temperature and for a given field orientation one should know not only the ratio $R(t)/R(1)$, but  the value of $R=H_{c2}/H_c$ itself, i.e., $R(T_c)$ or  the material parameters $\kappa_a$ and $\kappa_\parallel$.


  \begin{figure}[htb]
\begin{center}
 \includegraphics[width=7cm] {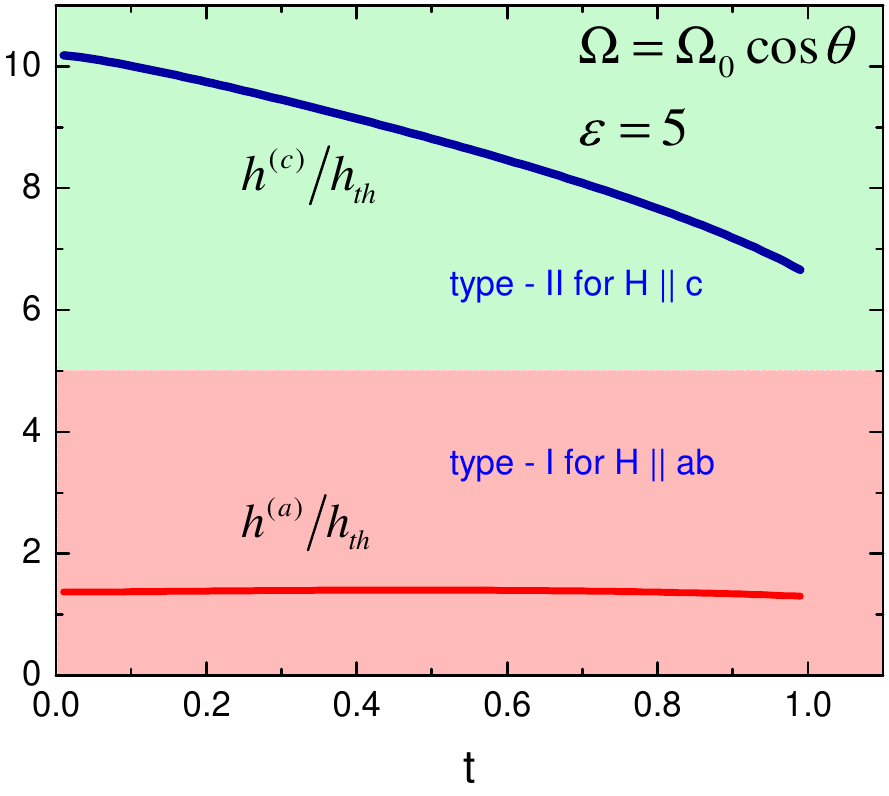}
\caption{(Color online) The  ratios $h^{(c)}/h_{th} $ and $h^{(a)}/h_{th} $ for the order parameter $\Omega=\Omega_0\cos\theta$ on a Fermi spheroid  with $\epsilon=5$   vs reduced temperature. For   $C\approx 0.2$, Eq.\,(\ref{const}), this corresponds to
$[H_{c2}^{(c)}/H_c]_{T_c}=\sqrt{2} \kappa_a(T_c) \approx 1.32$ and  $[H_{c2}^{(a)}/H_c]_{T_c}=\sqrt{2} \kappa_{\parallel}(T_c) \approx 0.26$.  A  hypothetic  superconductor   with such characteristics  is of type-II   in magnetic field along the $c$ axis and  of type-I in fields along the $ab$ plane.
 }
\label{f5}
\end{center}
\end{figure}

\begin{figure}[htb]
\begin{center}
 \includegraphics[width=8.cm] {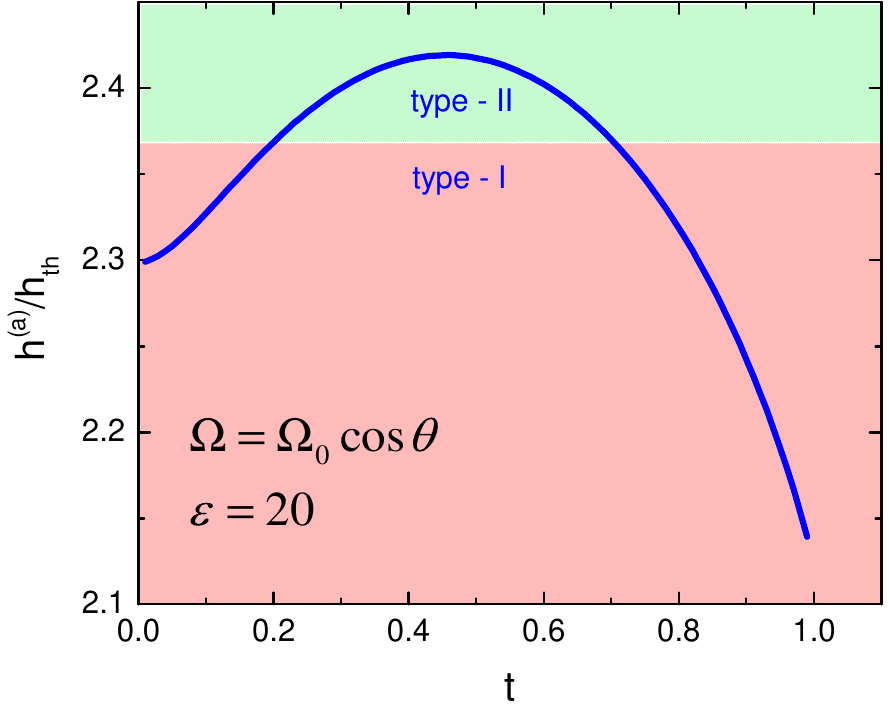}
\caption{(Color online) The ratio $h^{(a)} /h_{th} = (H_{c2}^{(a)} /H_c)/C $ vs $t$ for $\epsilon=20$, $\Omega=\Omega_0 \cos\theta$.  The boundary between the type-II and type-I corresponds to the constant of Eq.\,(\ref{const}) $C\approx 0.42$.
 }
\label{f6}
\end{center}
\end{figure}

 Other interesting possibilities are depicted in Figs.\,\ref{f5} and \ref{f6}. In Fig\,\ref{f5}
the  ratios $h^{(c)}/h_{th} $ and $h^{(a)}/h_{th} $ for the order parameter $\Omega=\Omega_0\cos\theta$ on a Fermi spheroid  with $\epsilon=5$ are plotted vs temperature. According to Eq.\,(\ref{Hc2(t)/Hc(t)}) to get the ratio of actual $H_{c2}^{(c)}/H_c$ one has to multiply $h^{(c)}/h_{th}$ by a material specific constant $C$ which is roughly  estimated as
 \begin{eqnarray}
C= \frac{  \phi_0 T_c }{\hbar^2v_F^2    \sqrt{  N(0)}} \approx 0.1\, T_c(K)  \,,
  \label{const}
   \end{eqnarray}
where we took   $v_F\approx 10^8\,$cm/s and $N(0)\approx 10^{33}\,$1/erg\,cm$^3$. If, for example,   $C\approx  0.2$,  the ratio  $H_{c2}^{(c)}/H_c > 1$ according to  Fig.\,\ref{f5}, while $H_{c2}^{(a)}/H_c < 1$ for all temperatures.   In other words, in this hypothetic situation the material is of type-II in fields along the $c$ axis and of type-I in fields perpendicular to $c$.

The lower panel of Fig.\,\ref{f4} shows that when the field is in the $ab$ plane  the ratio $R^{(a)}=H_{c2}^{(a)}/H_c$ is a non-monotonic function of $t$ for an oblate Fermi spheroid. The source of this behavior is  in the fact that $R^{(a)}=\gamma_H R^{(c)}$ and, as shown in Ref.\,\onlinecite{arXiv}, for the order parameter $\propto\cos\theta$, $\gamma_H$ increases on warming. To verify that this behavior is not accidental we have calculated this ratio for $\epsilon\gg 1$ which corresponds to nearly one-dimensional situation, Fig.\,\ref{f6}.
This example shows that, in principle,  situations are possible for which   two transitions from type-I to type-II and back happen with changing temperature.

Whether or not such  scenarios are realistic remains to be seen. It is known that clean elemental metals have rather small $\kappa_{GL}$. Usually, new superconducting compounds are of a strong type-II with $\lambda/\xi\gg 1$. It is not excluded, however, that an anisotropic material with small $\kappa$ will be discovered in future.

\section{Discussion}

We have shown that the criterion for the type of superconductivity based on the value of $\lambda/\xi$ established for the GL domain near $T_c$  cannot be used at arbitrary temperatures.
The criterion based on the inequality $H_{c1}<H_c$  cannot be used  because there is apparently no straightforward way to calculate  the line energy of a single vortex at arbitrary $T$ which is directly related to   $H_{c1}$. On the other hand, both the upper critical field $H_{c2}$ and the thermodynamic one, $H_c$, can be evaluated exactly at any $T$ for any anisotropy.  This qualifies the inequality $H_{c2}(T)>H_c(T)$ as an exact criterion for the type-II superconductivity.

While  evaluating   $R=H_{c2}/H_c$   within the microscopic theory, we do not observe any peculiarities near $R(T_c)=1$ of the sort   discussed in literature in the frame of extended GL equations for $\kappa_{GL}\approx 1/\sqrt{2}$, see Ref.\,\onlinecite{Luk'yanchuk} and references therein. Of course, if the curves of $H_{c2}(T)$ and $ H_c(T)$ cross at some $T^*<T_c$, the material should undergo transition from type-I to type-II or otherwise so that in the vicinity of $T^*$ one should take fluctuations   into account (along with the sample shape and  possibility of hysteresis), which are beyond  the mean-field BCS  theory. We, however, note that the argument   for existence of a broad region of the $HT$ phase diagram  well under $T_c$ with degenerate vortex configurations \cite{Bogomolnyi}  in materials with $\kappa_{GL}\approx 1/\sqrt{2}$    is essentially mean-field as well.\cite{Luk'yanchuk}

 Clearly,  models based on extended GL  functional are perfectly legitimate for systems described by this functional,  provided this functional is considered as exact.
  However, for superconductors, the GL theory is an approximation which holds for $T\to T_c$ within certain accuracy.  To study superconductors behavior  in extended $T$ domain, one should use, if possible, the microscopic theory instead of considering exact consequences of an approximate GL functional. As far as relative values of $H_{c2}$ and $H_c$ are concerned,  this has been done    for isotropic bulk materials by Eilenberger, \cite{Eil} who found that even if $H_{c2}(T_c)= H_c(T_c)$ or $\kappa_{GL}=1/\sqrt{2}$, $H_{c2}$ increases faster than $H_c$ with reducing $T$  ($dH_{c2}/dT|_{T_c}> dH_{c}/dT|_{T_c}$). Hence, there is no finite region of temperatures near $T_c$ where $H_{c2}(T)= H_c(T)$. This in fact contradicts the claim of Ref.\,\onlinecite{Luk'yanchuk} that such a region does exist.
 For anisotropic one-band superconductors considered here, the microscopic approach also does not give an indication of  peculiarities of the system properties for $R(T_c)=1$   (such as degeneracy of different vortex configurations \cite{Bogomolnyi}  in a broad region of the $HT$ phase diagram).

\appendix

\section{Averaging over Fermi spheroids}

Consider an  uniaxial superconductor with the electronic spectrum
\begin{equation}
E(\bm k)=\hbar^2\left(\frac{k_x^2+k_y^2}{2m_{ab}}+\frac{k_z^2}{2m_c}\right)\,,
\label{spectr}
\end{equation}
so that the Fermi surface is a spheroid  with $z$ being the symmetry axis.
In spherical coordinates $(k,\theta,\phi)$ we have
\begin{equation}
E(\bm k)=\frac{\hbar^2k^2}{2m_{ab}}\left(
\sin^2\theta+\frac{m_{ab}}{m_c}\cos^2\theta
\right)=\frac{\hbar^2k^2}{2m_{ab}}\Gamma(\theta),\qquad
\end{equation}
so that
\begin{equation}
k_F^2(\theta)=\frac{ 2m_{ab} E_F }{\hbar^2 \Gamma(\theta)} \,.
\label{kF}
\end{equation}

 \begin{figure}[b]
\begin{center}
 \includegraphics[width=8.cm] {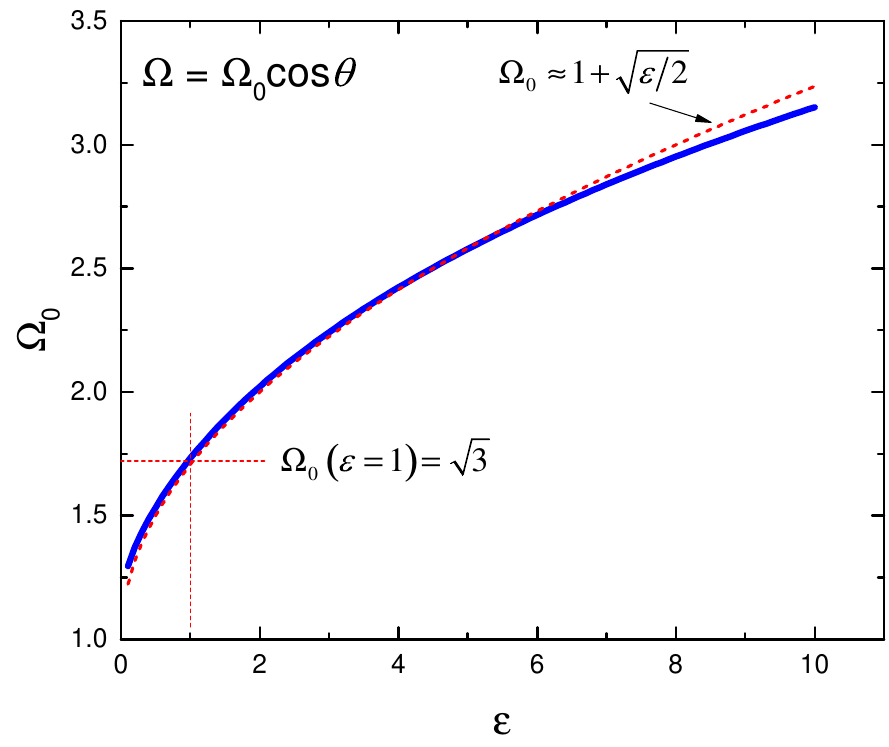}
\caption{(Color online) The normalization constant $\Omega_0$ for the order parameter $\Omega=\Omega_0\cos\theta$ as a function of the Fermi surface shape parameter $\epsilon$. The dashed curve is a convenient approximation to $\Omega_0(\epsilon)$.
 }
\label{f7}
\end{center}
\end{figure}

The Fermi velocity is   $\bm v(\bm k)=\bm\nabla_{\bm k}E(\bm k)$,
with the derivatives taken at $\bm k=\bm k_F$:
\begin{eqnarray}
v_x&=&\frac{v_{ab} \sin\theta\cos\phi}{\sqrt{\Gamma(\theta)}}, \,\,\,
v_y=\frac{v_{ab}\sin\theta\sin\phi}{\sqrt{\Gamma(\theta)}},\,\,\,\nonumber \\
 v_z&=&\epsilon \frac{v_{ab}\cos\theta}{\sqrt{\Gamma(\theta)}},\quad
 \epsilon=\frac{m_{ab}}{m_c}\,,   \quad v_{ab}= \sqrt{ \frac{2E_F}{m_{ab}} }\,. \quad
 \label{velocities}
\end{eqnarray}
The  value of the local Fermi velocity,  $v =(v_x^2+v_y^2+v_z^2)^{1/2}$, is given by
\begin{equation}
v= v_{ab}\sqrt{\frac{
 \sin^2\theta+\epsilon^2 \cos^2\theta
}{\sin^2\theta+\epsilon\cos^2\theta}}=v_{ab}\sqrt{\frac{\Gamma_1(\theta)}{\Gamma(\theta)}} \,.
\end{equation}
The density of states   is:
\begin{eqnarray}
N(0)=\int\frac{\hbar^2d^2\bm k_F}{(2\pi\hbar)^3v} = \frac{m^2_{ab} v_{ab}}{2\pi^2\hbar^3}
\int\frac{d\Omega}{4\pi \sqrt{\Gamma(\theta)\Gamma_1(\theta)}} ,\qquad
\label{eqDOS}
\end{eqnarray}
where the integration  is  over the
solid angle $d\Omega=\sin\theta\;d\theta\, d\phi$.

 The Fermi surface average of a function $A( \theta,\phi)$   is
 \begin{eqnarray}
\langle A( \theta,\phi) \rangle
 =\frac{1}{D}
 \int\frac{d\Omega\,A( \theta,\phi)}{4\pi  \sqrt{\Gamma(\theta)\Gamma_1(\theta)}}\,,\qquad\qquad\label{<A>}\\
D = \int\frac{d\Omega}
{ 4\pi \sqrt{\Gamma(\theta,\epsilon)\Gamma_1(\theta,\epsilon)}} =\frac{F(\cos^{-1}\sqrt{\epsilon},1+\epsilon)}{\sqrt{1-\epsilon}}  \qquad
\label{D_el}
\end{eqnarray}
where $F $ is an Incomplete Elliptic Integral of the first kind. If  $A$ depends only on the polar angle $\theta$, one can employ $u=\cos \theta$:
\begin{eqnarray}
\langle A( \theta ) \rangle
=\frac{1}{D(\epsilon)}
 \int_0^1\frac{du\,A(u)}{ \sqrt{\Gamma(u,\epsilon)\Gamma_1(u,\epsilon)}}\,,\qquad\qquad\label{<A1>}\\
\Gamma =  1+(\epsilon-1)u^2\,,\qquad \Gamma_1  = 1+(\epsilon^2-1)u^2 \,.\qquad
\label{average1}
\end{eqnarray}
It is useful to have a relation between $v_{ab}=\sqrt{2E_F/m_{ab}}$ and $v_0$ of  Eq.\,(\ref{mu_c}) for a one-band situation:
\begin{eqnarray}
v_{ab}^3= D (\epsilon)\, v_0^3 \,.
 \label{v0-vpar}
\end{eqnarray}

As an example we show in Fig.\,\ref{f7} how the averaging over  Fermi spheroids affects the normalization constant $\Omega_0$ for the order parameter of the form $\Omega=\Omega_0\cos\theta$.


\bibstyle{apsrev4-1}


\begin{thebibliography}{99}

\bibitem{GL}V. L. Ginzburg and L. D. Landau, Zh. Eksperiment. i Teor. Fiz {\bf 20}, 1064 (1950).

\bibitem{Abrik} A. A. Abrikosov, Sov. Phys. JETP {\bf 5}, 1174 (1957).

\bibitem{ChiaRenHu} Chia-Ren Hu, \prb {\bf 6}, 1756 (1972).

\bibitem{Eil} G. Eilenberger, Phys. Rev.  {\bf  153}, 584 (1967).

\bibitem{Buzdin}A. Buzdin and A. Simonov, JETP Lett. {\bf 50},  325  (1989).

 \bibitem {K2002} V.G. Kogan,   \prb {\bf  66}, 020509 (2002).

 \bibitem{PK-ROPP}R. Prozorov and V.G. Kogan,   Rep.  Progr.   Phys.  {\bf 74}, 124505 (2011).

  \bibitem{KP-ROPP}V.G. Kogan and R. Prozorov,      Rep.  Progr.  Phys.  {\bf 75}, 114502 (2012).


       \bibitem{E}G. Eilenberger, Z. Phys. {\bf  214}, 195 (1968).

\bibitem{Kad} D. Markowitz and L.P. Kadanoff, Phys. Rev.  {\bf  131},
363 (1963).

\bibitem{Nagi}G. Haran, J. Taylor, and A. D. S. Nagi, \prb {\bf 55}, 11778
(1997).

\bibitem{Openov}L. A. Openov, \prb {\bf 69}, 224516 (2004).



 \bibitem{MMK} P. Miranovi\' c, K. Machida,  V. G. Kogan,  J. Phys. Soc. of Japan  {\bf 72}, No.2,  221 (2003)





\bibitem{theory-nodes} V. Mishra, S. Graser, and P. J. Hirschfeld, \prb {\bf 84}, 014524 (2011).


\bibitem{arXiv1}
    R. S. Gonnelli, D. Daghero, M. Tortello, G. A. Ummarino, Z. Bukowski, J. Karpinski, P. G. Reuvekamp, R. K. Kremer, G. Profeta, K. Suzuki, K. Kuroki, arXiv:1406.5623.


\bibitem{Feng}Y. Zhang, Z. R. Ye, Q. Q. Ge, F. Chen, Juan Jiang, M. Xu, B. P. Xie, D. L. Feng,     Nature Physics, doi:10.1038/nphys2248 (2012).


 \bibitem{HW}E. Helfand, N.R. Werthamer, Phys. Rev. {\bf 147}, 288 (1966).

\bibitem{Gorkov}L. P. Gor'kov, Sov. Phys. JETP,

\bibitem{Parks} A. Fetter and P. Hohenberg in {\it Superconductivity}, ed R. D. Parks, Marcel Dekker, New York, v. 2,  1969.


\bibitem{Gork}L.P. Gor'kov and T.K. Melik-Barkhudarov, Soviet Phys. JETP,
{\bf  18}, 1031 (1964).

\bibitem{arXiv}V.G. Kogan and R. Prozorov, arXiv:1405.2359.





\bibitem{Carrington} J.D. Fletcher, A. Carrington, O.J. Taylor, S.M.
Kazakov, J. Karpinski, \prl {\bf  95}, 097005 (2005).




\bibitem{Luk'yanchuk}I. Luk'yanchuk, \prb {\bf 63}, 174504 (2001).



\bibitem{Bogomolnyi}E. B. Bogomolnyi,  Sov. J. Nucl. Phys.
 {\bf 24}, 449 (1976); E. B. Bogomolnyi and A. I. Vainstein, ibid.  {\bf 23}, 588 (1976).

\end{thebibliography}
\end{document}